\documentclass[twocolumn,showpacs,preprintnumbers,amsmath,amssymb,superscriptaddress,nofootinbib,pre]{revtex4}

\usepackage{dcolumn}
\usepackage{bm}
\usepackage{graphics}
\usepackage{graphicx}
\begin{document}

\title{General disagreement between the Geometrical Description of Dynamical In-stability  -using non affine parameterizations-  and traditional Tangent Dynamics}

\author{Eduardo Cuervo-Reyes}
\thanks{}
\email[Corresponding author: ]{cuervo@inorg.chem.ethz.ch }
\affiliation{LAC ETH-Zurich, Switzerland}
 
\date{\today}
\begin{abstract}
 In this paper, the general disagreement of the geometrical lyapunov exponent with lyapunov exponent from tangent dynamics is addressed. It is shown in a quite general way that the vector field of geodesic spread $\xi^k_G$ is not equivalent to the tangent dynamics vector $\xi^k_T$ if the parameterization is not affine and that  results regarding dynamical stability obtained in the geometrical framework can differ qualitatively from those in the tangent dynamics. It is also proved in a general way that in the case of Jacobi metric -frequently used non affine parameterization-, $\xi^k_G$ satisfies differential equations which differ from the equations of the tangent dynamics in terms that produce parametric resonance, therefore, positive exponents for systems in stable regimes.      
\end{abstract}


\maketitle

The definition of an appropriate indicator of dynamical (in)-stability for systems described by the hamiltonian
\begin{eqnarray}
H(\bm{q},\bm{p})=\frac{1}{2}a^{ij}p_ip_j+V(\bm{q}) && p_i=a_{ij}\dot{q}^j
\end{eqnarray}
\noindent has been the subject of many works during the last century. We can cite, those primary methods based on perturbation theory that appeared in the works of Poincare, usage of action variables, homoclinic intersections and qualitative analysis from the poincare surfaces sections PSS; or more quantitative and non-perturbative methods like the Lyapunov exponent\cite{Poincare,Arnold,Krylov,randy}. The latter, is a direct measure of the exponential divergence of close trajectories. It is computed as the limit
\begin{equation}
\lambda=\lim_{t\rightarrow\infty}\frac{1}{t}\ln(\frac{||\bm{q}(t,\tau_2)-\bm{q}(t,\tau_1)||}{||\bm{q}(0,\tau_2)-\bm{q}(0,\tau_1)||})
\end{equation} 
\noindent where $\tau$ is a set of parameters which determines the initial conditions for each trajectory\footnote{In order to simplify the notation, here and in the following I omit the evident dependence of the exponent on the initial conditions and I write $\lambda$ instead of $\lambda(\tau)$}. In this way $\lambda$ quantifies the exponent of divergence of two trajectories that have initial conditions $\tau_1$ and $\tau_2$ close to each other. This indicator, although very intuitive, clearly fails for bounded systems, where trajectories can evolve very different but the distance $||\bm{q}(t,\tau_2)-\bm{q}(t,\tau_1)||$ will never be greater than the size of the system.
An alternative that was proposed to solve this problem was the tangent dynamics. This method, calculate a similar exponent but based on a different measure. The distance  $||\bm{q}(t,\tau_2)-\bm{q}(t,\tau_1)||$ is replaced by the norm of the solution $\Delta \bm{q}$ of the linearized variation of the equations of motion\cite{Reinel}.
\begin{subequations}
\begin{eqnarray}
\Delta {\dot p}^i=-\left(\Delta q^j\frac{\partial}{\partial q^j}+\Delta p_j\frac{\partial}{\partial p_j}\right)\frac{\partial H(\bm{q},\bm{p})}{\partial q^i}\\
\Delta {\dot q}^i=\left(\Delta q^j\frac{\partial}{\partial q^j}+\Delta p_j\frac{\partial}{\partial p_j}\right)\frac{\partial H(\bm{q},\bm{p})}{\partial p^i}
\end{eqnarray}
\end{subequations}
It is equivalent to follow the evolution of 
\begin{equation}
\bm{\xi}_T=\lim_{\tau_2\rightarrow\tau_1}\frac{\bm{q}(t,\tau_2)-\bm{q}(t,\tau_1)}{\tau_2-\tau_1}=\left(\frac{\partial\bm{q}}{\partial\tau}\right)_t\label{tangent}
\end{equation} 
\noindent a directional derivative which is taken at constant  $t$\footnote{Directional in the sense that the parameter space has more that one dimension. Generally, if the system has $N$ degrees of freedom, $2N$ parameters are needed to determine the trajectory.}.

 Although the system might be bounded, the tangent dynamics accumulate the local gradients of divergence along the trajectory and it has proved to give a good measure of how sensitive a trajectory is with respect to variation of initial conditions. 

In the first half of the 20$th$ Eisenhart, inspired by general relativity theory, proposed a geometrical description of classical dynamics for systems which can be described by the least action principle\cite{Eisenhart}. Geometrization of dynamics represents the motion (actual trajectories) as a geodesic on a manifold with a suitable metric $g=g_{\mu\nu}dq^{\mu}\otimes dq^{\nu}$. For this purpose he introduced a metric that has been known after him as Eisenhart metric, whose arc length is given by 
\begin{equation}
ds^2=-2V(\bm{q})(dq^0)^2+a_{ij}dq^{i}dq^{j}+2dq^{0}dq^{N+1}
\end{equation}
Here $q^0=t$ and, from the integration of geodesics equation, the extra coordinate is
\begin{equation}
q^{N+1}=\frac{\kappa^2}{2}t+C_0-\int_0^t {\cal L}dt'
\end{equation}
\noindent Physical motions satisfy an {\it affine parametrization} $ds^2=\kappa^2 dt^2$ ($\kappa$ is a real arbitrary constant).

It has been proposed that from a geometrical framework, stability analysis are also possible, by means of the geometrical lyapunov exponent\cite{Pettini1}.
\begin{equation}
\lambda_G=\lim_{s\rightarrow\infty}\frac{1}{s}\ln(\frac{||\bm{\xi}_G(s)||}{||\bm{\xi}_G(0)||})
\end{equation}
\noindent Here
 \begin{equation}
\xi_G^i(s)=\left[\frac{\partial q^i(\tau,s)}{\partial\tau}\right]_s \label{xi}
\end{equation} 
\noindent and it is called the vector field of geodesic spread. It is a directional derivative of geodesics at fixed arc length and evolves according to the Jacobi-Levi-Civita equation  
\begin{equation}
\frac{D^2\xi^i}{ds^2}+R^i_{jkm}\frac{dq^j}{ds}\xi^k\frac{dq^m}{ds}=0\label{JLC}
\end{equation}
\noindent where $\frac{D}{ds}$ and $R^i_{jkm}$ stand respectively for the covariant derivative along geodesics and the Riemann-Christoffel curvature tensor\cite{Levi}. Eq.(\ref{JLC}), after opening the covariant derivative and the curvature tensor reads
\begin{equation}
\frac{d^2\xi^k}{ds^2}+2\Gamma^k_{lj}\frac{dq^l}{ds}\frac{d\xi^j}{ds}+\Gamma^k_{lm,j}\frac{dq^l}{ds}\frac{dq^m}{ds}\xi^j=0\label{JLCopen}
\end{equation}
\noindent where $\Gamma^k_{lj}=\frac{1}{2}g^{km}\left(g_{lm,j}+g_{mj,l}-g_{lj,m}\right)$.
It has been shown that for the Eisenhart metric the spacial components of JLC equations are equivalent to the equations of tangent dynamics.

Several works in the last decade have been dedicated to show that another metric can be used, in particular, the kinetic energy metric (also called Jacobi metric)\cite{Pettini2} 
\begin{equation}
(g_J)_{ij}\equiv 2[E-V(\bm{q})]a_{ij}(\bm{q})\label{Jacg}
\end{equation} 
By applying the two mentioned metrics to some unstable systems it has been argued the equivalence of the results\cite{Pettini7,Reinel}, although in some cases a strange suppression of chaos with increasing number of non-separable degrees of freedom has been observed using Jacobi metric\cite{Pettini2}. The fact that for many hamiltonian systems the curvature in the Jacobi framework is mostly positive has been used to support the idea that the main source of instability is the parametric resonance rather than the hiperbolicity of the potential surface\cite{Reinel,Pettini2,Pettini3}.
There are some reasons already declared in previous works pointing at the inadequacy of the Jacobi metric. One reason is that it becomes singular at the boundaries where the kinetic energy is zero. Thus, there are geodesics on the boundaries that do not correspond to physical motions\cite{szyd1,szyd2}. However, some authors consider that for many degrees of freedom, where the probability of reaching the boundaries is practically zero, this metric gives good physical results, but it has not been proved. Some stimulating results regarding geometrization in the thermodynamic limit have been obtained just within Eisenhart metric \cite{Pettini4,Pettini5}. A general proof of the equivalence of the geometrical approach and the tangent dynamics, is still missing; while Cuervo and Movassagh\cite{EdRam} have already shown the appearance of a positive lyapunov exponent in a trivially stable system when Jacobi metric is used.   

Here I will show briefly that the geometrical  measure (\ref{xi}) generally does not give results equivalent to those from the tangent dynamics when parameterization is not affine and I will explicitly show how the jacobi metric fails. 

Let $q^i$ be the coordinates used to describe a system, $t$ the time and $\tau$ the set of parameters that determine the initial conditions. The motions (trajectories) $q^i(t,\tau)$ will depend of time and $\tau$. The arc length is a functional of the path; on extremal paths the arc length is a function  of time and $\tau$. On geodesics, coordinates and arc length variations are
\begin{subequations}
\begin{eqnarray}
ds=\left(\frac{\partial s}{\partial\tau}\right)_t d\tau+\left(\frac{\partial s}{\partial t}\right)_{\tau}dt\label{vars}\\
dq^i=\left(\frac{\partial q^i}{\partial\tau}\right)_td\tau+\left(\frac{\partial q^i}{\partial t}\right)_\tau dt\label{varq}
\end{eqnarray}
\end{subequations}
\noindent where $\left(\frac{\partial s}{\partial\tau}\right)_t$ is the variation of the arc length  when one moves from one trajectory to another, at fixed time. $(\frac{\partial }{\partial\tau })_t$ is just a short form of $({\bf n}\cdot\nabla_\tau)_t$, a directional derivative on the initial conditions taken at fixed time. From eqs. (\ref{vars}) and (\ref{varq}) 
\begin{subequations}
\begin{eqnarray}
0=\left(\frac{\partial s}{\partial\tau}\right)_t+\left(\frac{\partial s}{\partial t}\right)_{\tau}\left(\frac{dt}{d\tau}\right)_s\label{nulds}\\
\left(\frac{\partial q^i}{\partial\tau}\right)_s=\left(\frac{\partial q^i}{\partial\tau}\right)_t+\left(\frac{\partial q^i}{\partial t}\right)_\tau \left(\frac{dt}{d\tau}\right)_s \label{varqdtau0}
\end{eqnarray}
\end{subequations}
\noindent  and using the definitions Eqs. (\ref{tangent}) and (\ref{xi}), one arrives to 
\begin{equation}
\xi_G^i(t,\tau)=\xi_T^i(t,\tau)-\dot{q}^i\left(\frac{\partial  s}{\partial\tau}\right)_t\left(\frac{\partial s}{\partial t}\right)_{\tau}^{-1}\label{main0}
\end{equation}
 
Eq.(\ref{main0}) relates in a compact way, the definitions  of the divergence vector field obtained from tangent dynamics with that obtained from any geometrical approach. 

A consequence of Eq.(\ref{main0}) is that geometrical $\xi_G$ and tangent dynamics  $\xi_T$ measures will differ unless $\left(\frac{\partial s}{\partial\tau}\right)_t$ vanish. This is the case of affine parameterizations, where the arc length ($ds^2=\kappa^2dt^2$) depends just of the time elapsed. This is consistent with the fact that in Eisenhart metric $\kappa$ is an arbitrary positive constant independent of the trajectory and results regarding stability are always equivalent to those obtained from the tangent dynamics. In general cases, the directional derivative $\left(\frac{\partial s}{\partial\tau}\right)_t$ might have singularities or be ill-defined. 

It is evident that if $\left(\frac{\partial s}{\partial t}\right)_{\tau}$ has zeros along the trajectory the geometrical measure will diverge. This is the case of Jacobi metric ($\left(\frac{\partial s}{\partial t}\right)_{\tau}=2T$) when the kinetic energy can be zero (very common in low dimensional systems). However, systems with more degrees of freedom, where  zeros in the kinetic energy are not likely to happen, require a further analysis. For it, one can look at the differential equations satisfied by $\xi^k_J$
and $\xi^k_T$. For the Tangent dynamics the equations of motion are
\begin{widetext}
\begin{subequations}
\begin{eqnarray}
\ddot{\xi}_T^n&=&-\left(a^{nk}V_{,kl}+a^{nk}_{,l}V_{,k}+t^n_{ij,l}\dot{q}^i\dot{q}^j\right)\xi_T^l-2t^n_{ij}\dot{\xi_T}^i\dot{q}^j \\
t^n_{ij}&=&\frac{1}{2}a^{nk}\left(a_{ki,j}+a_{kj,i}-a_{ij,k}\right)
\end{eqnarray}
\end{subequations}
\noindent while for the Jacobi measure, after substituting  $\Gamma^k_{lj}=t^k_{lj}+\frac{1}{T}\left(a^{km}a_{lj}V_{,m}-\delta^k_{j}V_{,l}-\delta^k_{l}V_{,j}\right)$ in (\ref{JLCopen}) one obtains
\begin{eqnarray}
\ddot{\xi}_J^n&=&-\left(a^{nk}V_{,kl}+a^{nk}_{,l}V_{,k}+t^n_{ij,l}\dot{q}^i\dot{q}^j\right)\xi_J^l-2t^n_{ij}\dot{\xi_J}^i\dot{q}^j \nonumber\\
&&-\frac{1}{T}\left( a^{nm}V_{,m}\lbrace a_{ij}\dot{q}^i\dot{\xi}^j_J+V_{,l}\xi^l+\frac{1}{2}a_{ij,l}\dot{q}^i\dot{q}^j\xi^l_J\rbrace-\dot{q}^n\lbrace V_{,il}\dot{q}^i\xi^l_J+V_{,j}\dot{\xi}^j_J+\frac{1}{T}V_{,i}\dot{q}^iV_{,l}\xi^l_J\rbrace\right)\label{Jacdyn}\nonumber\\
&&
\end{eqnarray}
\end{widetext}
The equation for $\xi_J$ contains the same terms that the one for $\xi_T$ plus  other terms which have been written in the second line. The first conclusion that can be drown from it is the already known appearance of divergent terms when kinetic energy is zero. In addition these terms, as will be shown immediately, introduce non physical instabilities also for trajectories where kinetic energy never vanishes. 

 Many of the systems of interest are bounded. The phase space has a finite volume. For energies close to a minimum of the potential, the motion will be restricted to small oscillations around an equilibrium point. In this case, harmonic approximation for the potential energy is valid and  systems are dynamically stable with periodic trajectories. Since the stability analysis should not depend of the variables  used, let expand for simplicity, the potential around a minimum in cartecian coordinates. So, $a_{ij}$ are constant, $t^n_{ij}\equiv 0$.
 
 Tangent dynamics equations become \begin{eqnarray}
\ddot{\xi}_T^n&=&-a^{nk}V_{,kl}\xi^l_T
\end{eqnarray}
For up to second order expansion in the potential $V({\bf q})$ these equations give oscillatory solutions for $\xi_T$ leading to vanishing exponent. When the system is more excited above the minimum energy, terms of order higher than two in coordinate dependence of the potential must be considered.  Here motion will be quasi-periodic and this regime corresponds to the onset of chaos. $V_{,kl}$ (the frequencies of the tangent dynamics) will be ${\bf q}$-dependent (time dependent, periodic)\footnote{Expansion of $V$ up to $n$-order ($n\ge 2$) in ${\bf q}$ gives $V_{,kl}$ of order $n-2$}. Since $V_{,kl}$ oscillate at harmonics of the frequencies of the system they produce parametric resonance in the tangent dynamics and therefore positive exponents. For very high energies, systems start filling the phase space, trajectories are no longer periodic but unstable due to hyperbolic points between the minimums of the potential. Thus, parametric resonance in $\xi$ dynamics is relevant at the on-set of chaos in hamiltonian systems.

The picture is very different in the Jacobi framework. The extra terms in the equations of motion for  $\xi_J$, (see bellow in cartecian coordinates eq.(\ref{Jacdyncart}))
\begin{widetext}
\begin{eqnarray}
\ddot{\xi}_J^n&=&-a^{nk}V_{,kl}\xi_J^l-\frac{1}{T}\left( a^{nm}V_{,m}\lbrace a_{ij}\dot{q}^i\dot{\xi}^j_J+V_{,l}\xi^l\rbrace-\dot{q}^n\lbrace V_{,il}\dot{q}^i\xi^l_J+V_{,j}\dot{\xi}^j_J+\frac{1}{T}V_{,i}\dot{q}^iV_{,l}\xi^l_J\rbrace\right)\label{Jacdyncart}
\end{eqnarray}
\end{widetext}
\noindent contain ${\bf q}$-dependent terms also when the potential is quadratic in ${\bf q}$, which therefore oscillates with the frequencies of the system and its harmonics. For even potentials (it is always the case in the harmonic approximation), the kinetic energy oscillates with frequencies that are twice the frequencies of the system. So, the extra terms in eq.(\ref{Jacdyncart}) set the perfect scenario for parametric resonance, and it has nothing to do with non-linear behavior of the physical system. For systems at the onset of chaos, real parametric resonance appears from the first term of the rhs in Eq.(\ref{Jacdyncart}) and for higher energies the effect of the hyperbolic points will come into the game. Thus, it is not possible to discriminate the false exponential divergence from the physical one, when the exponent is computed from the evolution of $\xi_J$. Stable regimens can be shown as chaotic.  

Another point to stress is that Jacobi and Eisenhart metric are not related by a canonical transformation, they are defined in different spaces. The statement ``physics is independent of the frame of reference'', is then out of context since these coordinates must be defined on a same space and related by a canonical transformation. In addition, the independent variable (arc length) chosen in the Jacobi metric is not invariant with respect to canonical transformations.

In conclusion, whether the physical system is chaotic or not, and the nature of this instabilities (hyperbolic points or parametric resonance) are questions that can not be answered  by   Jacobi metric. It always gives positive exponents for hamiltonian bounded systems. Other non-affine parameterizations might have  similar drawbacks. In general, the vector field of geodesic spread and the tangent dynamic vector field are not equivalent when the parameterization with the arc length is not affine.

I wish to thank R. Movassagh, Reinel Sospedra-Alfonso and Luisberis Vel\'asquez for very valuable discussions. This project was funded by Swiss National Science Foundation.

\end{document}